\documentclass[superscriptaddress,nofootinbib,longbibliography]{revtex4-2}
\usepackage[latin9]{inputenc}
\usepackage{color}
\usepackage{verbatim}
\usepackage{units}
\usepackage{mathrsfs}
\usepackage{amsmath}
\usepackage{amssymb}
\usepackage{graphicx}
\usepackage[unicode=true,pdfusetitle,
 bookmarks=true,bookmarksnumbered=false,bookmarksopen=false,
 breaklinks=false,pdfborder={0 0 1},backref=false,colorlinks=true]
 {hyperref}
\hypersetup{
 citecolor=blue}

\makeatletter

\providecommand{\tabularnewline}{\\}

\usepackage{braket,bbm}

\makeatother

\begin{document}
\title{Spin-dependent dark matter-electron interactions}
\author{C.-P.~Liu}
\affiliation{Department of Physics, National Dong Hwa University, Shoufeng, Hualien
97401, Taiwan}
\affiliation{Physics Division, National Center for Theoretical Sciences, National
Taiwan University, Taipei 10617, Taiwan}
\author{Chih-Pan~Wu}
\affiliation{D\'{e}partement de physique, Universit\'{e} de Montr\'{e}al, Montr\'{e}al
H3C 3J7, Canada}
\author{Jiunn-Wei~Chen}
\affiliation{Department of Physics, Center for Theoretical Physics, and Leung Center
for Cosmology and Particle Astrophysics, National Taiwan University,
Taipei 10617, Taiwan}
\affiliation{Physics Division, National Center for Theoretical Sciences, National
Taiwan University, Taipei 10617, Taiwan}
\author{Hsin-Chang~Chi}
\affiliation{Department of Physics, National Dong Hwa University, Shoufeng, Hualien
97401, Taiwan}
\author{Mukesh~K.~Pandey}
\affiliation{Department of Physics, National Taiwan University, Taipei 10617, Taiwan}
\author{Lakhwinder~Singh}
\affiliation{Department of Physics, Central University of South Bihar, Gaya 824236,
India}
\affiliation{Institute of Physics, Academia Sinica, Taipei 11529, Taiwan}
\author{Henry~T.~Wong}
\affiliation{Institute of Physics, Academia Sinica, Taipei 11529, Taiwan}
\date{\today}
\begin{abstract}
Detectors with low thresholds for electron recoil open a new window
to direct searches of sub-GeV dark matter (DM) candidates. In the
past decade, many strong limits on DM-electron interactions have been
set, but most on the one which is spin-independent (SI) of both dark
matter and electron spins. In this work, we study DM-atom scattering
through a spin-dependent (SD) interaction at leading order (LO), using
well-benchmarked, state-of-the-art atomic many-body calculations.
Exclusion limits on the SD DM-electron cross section are derived with
data taken from experiments with xenon and germanium detectors at
leading sensitivities. In the DM mass range of 0.1--10 GeV, the best
limits set by the XENON1T experiment: $\sigma_{e}^{(\textrm{SD})}<10^{-41}-10^{-40}\,\textrm{cm}^{2}$,
are comparable to the ones drawn on DM-neutron and DM-proton at slightly
bigger DM masses. The detector's responses to the LO SD and SI interactions
are analyzed. In nonrelativistic limit, a constant ratio between them
leads to an indistinguishability of the SD and SI recoil energy spectra.
Relativistic calculations however show the scaling starts to break
down at a few hundreds of eV, where spin-orbit effects become sizable.
We discuss the prospects of disentangling the SI and SD components
in DM-electron interactions via spectral shape measurements, as well
as having spin-sensitive experimental signatures without SI background.
\end{abstract}
\maketitle

\section*{Introduction}

Direct searches of dark matter (DM) through its scattering with electrons
have being a rapidly growing field in the past decade. With low-threshold
capabilities of modern detectors in electron recoil (ER) and new ideas
inspired by theoretical studies, the coverage of DM mass, $m_{\chi}$,
has gradually been extended from sub-GeV towards increasingly lower
reach (for a comprehensive overview, see, e.g., Ref.~\citep{Battaglieri:2017aum}).
So far, most attention is given to the DM-electron interaction which
is spin-independent (SI) of both DM and electrons. Various experiments
have already set stringent exclusion limits on its cross section,
$\sigma_{e}^{(\textrm{SI})}$, in the mass range of MeV to GeV~\citep{Essig:2012yx,Essig:2017kqs,Agnes:2018oej,Crisler:2018gci,Agnese:2018col,Abramoff:2019dfb,Aprile:2019xxb,Aguilar-Arevalo:2019wdi,Arnaud:2020svb,Barak:2020fql,Amaral:2020ryn,Cheng:2021fqb}.
The current best limit above 30 MeV is set by XENON1T for their huge
exposure mass and time~\citep{Aprile:2019xxb,Pandey:2018esq}; in
the range of $1-30$ MeV, several experiments capitalizing the condensed
phases of materials such as semiconductor silicon~\citep{Aguilar-Arevalo:2019wdi,Barak:2020fql}
and germanium~\citep{Amaral:2020ryn,Arnaud:2020svb} show potential
improvements upon future scale-up. But, what about the spin-dependent
(SD) interactions?

While there can be numerous models to motivate studies of SD DM-electron
interactions from top down, a more straightforward, bottom-up approach
is through nonrelativistic (NR) effective field theory~\citep{Fan:2010gt,Fitzpatrick:2012ix,DelNobile:2013sia,Anand:2013yka,Anand:2014kea,DelNobile:2018dfg,Catena:2019hzw,Trickle:2020oki}.
As most DM is believed to be cold, its velocity $v_{\chi}\sim10^{-3}$
provides a natural parameter by which the DM interactions can be formulated
as a series expansion. At leading order (LO), i.e., $v_{\chi}^{0}=1$,
there are only two terms: $\mathbbm{1}_{\chi}\cdot\mathbbm{1}_{e}$
(SI) and $\vec{S}_{\chi}\cdot\vec{S}_{e}$ (SD), where $\mathbbm{1}$
and $\vec{S}$ are unity and spin operators, respectively. It is when
matching these NR effective operators to their underlying theories
that the generality and complementarity of such consideration shows
its full power. For example, at the relativistic level, the SI term
can come from either a scalar DM with a scalar-scalar coupling, or
a fermionic DM with a scalar-scalar or vector-vector coupling to electrons;
the SD term can come from a fermionic DM with an pseudovector-pseudovector
or tensor-tensor coupling. As a result, null observations of them
constrain different parts of the broad DM parameter space. On the
other hand, an observation of the SD term would not only be a discovery
of DM and a new interaction, but also exclude all scalar DM scenarios
-- this adds a stronger motivation to search for it. 

The LO SD DM-nucleon interactions have been studied, though not as
extensively as its SI counterpart, with good recent progress ~\citep{Akerib:2017kat,Agnese:2017jvy,Jiang:2018pic,Xia:2018qgs,Armengaud:2019kfj,Aprile:2019dbj,Amole:2019fdf,Abdelhameed:2020qen}.
The best limits on the contact interactions with a neutron $(n)$
and a proton ($p$), in terms of total cross sections, are $\sigma_{n}^{(\textrm{SD})}<6.2\times10^{-42}\,\textrm{cm}^{2}$
with $m_{\chi}=30\,\textrm{GeV}$ by XENON1T~\citep{Aprile:2019dbj},
and $\sigma_{p}^{(\textrm{SD})}<3.2\times10^{-41}\,\textrm{cm}^{2}$
with $m_{\chi}=25\,\textrm{GeV}$ by PICO~\citep{Amole:2019fdf},
respectively. Compared with the best limits of order $10^{-46}\,\textrm{cm}^{2}$
on the SI interactions, the huge order-of-magnitude difference is
easily understood from the (almost) coherent contributions of $A\sim100$
nucleons for the SI and a single unpaired nucleon for the SD case,
respectively, in scattering amplitudes. On the other hand, because
a typical atom is of size $\sim\mathring{\textrm{A}}$, a cold DM
particle can not induce substantial coherent scattering with atomic
electrons unless it is as light as $m_{\chi}\lesssim\textrm{MeV}$.
Therefore, in direct searches of DM in the MeV-GeV mass range, one
can anticipate constraints of similar orders on both LO SD and SI
DM-electron interactions, for the latter has no coherent enhancement
in scattering cross sections.

In this work, we study DM-atom scattering through the SD DM-electron
interaction at leading order, using well-benchmarked, state-of-the-art
atomic many-body approaches. Exclusion limits are derived from various
data of xenon and germanium detectors. The limit on DM-electron cross
section $\sigma_{e}^{(\textrm{SD})}<10^{-41}-10^{-40}\,\textrm{cm}^{2}$
with $m_{\chi}=0.1-10\,\textrm{GeV}$, set by XENON1T data, is comparable
to the best ones on $\sigma_{n,p}^{(\textrm{SD})}$ mentioned above
at slightly bigger $m_{\chi}$. Special attention is on the difference
in detector responses to the SI and SD interactions, with spin-orbit
effects being found to be the deciding factor. We discuss in the end
several strategies for sub-GeV DM searches that can disentangle the
SD and SI DM-electron interactions.

\section*{Formalism}

At leading order (LO), the effective SD interaction between DM ($\chi$)
and an electron ($e$) is 
\begin{equation}
\mathscr{L}_{\textrm{SD}}^{(\textrm{LO})}=(c_{4}+d_{4}/q^{2})\left(\chi^{\dagger}\vec{S}_{\chi}\chi\right)\cdot\left(e^{\dagger}\vec{S}_{e}e\right)\,,\label{eq:L_SD}
\end{equation}
where the coupling constants $c_{4}$ and $d_{4}$ (following the
convention of Refs.~\citep{Fan:2010gt,Fitzpatrick:2012ix}) are for
the short- and long-range interactions, respectively; and $q$ the
magnitude of the three-momentum transfer $\vec{q}$. 

The unpolarized differential scattering cross section in the laboratory
frame can be derived straightforwardly (see, e.g., Ref.~\citep{Chen:2015pha}),
and we focus only on the ionization processes that yield ER signals
\begin{equation}
\frac{d\sigma}{dT}^{(\textrm{ion})}=\frac{1}{2\pi v_{\chi}^{2}}\int qdq\,(\bar{c}_{4}+\bar{d}_{4}/q^{2})^{2}R_{\textrm{SD}}^{(\textrm{ion})}(T,q).\label{eq:dcs^SD}
\end{equation}
Note that for later convenience we redefine the coupling constants
$(\bar{c}_{4},\bar{d}_{4})=\sqrt{\bar{s}_{\chi}/4}(c_{4},d_{4})$
by absorbing (i) the spin factor $\bar{s}_{\chi}=s_{\chi}(s_{\chi}+1)/3$,
which results from an average of the initial, and a sum of the final
DM spin states, with $s_{\chi}=1/2$ or $1$ applying to the case
of a fermionic or vector DM particle, and (ii) the square of the electron
spin $s_{e}=1/2$. The SD response function 
\begin{align}
R_{\textrm{SD}}^{(\textrm{ion})}(T,q)= & \sum_{\mathscr{F}}\overline{\sum_{\mathscr{I}}}\sum_{j}\,|\braket{\mathscr{F}|\sum_{i=1}^{Z}e^{i\vec{q}.\vec{r}_{i}}\sigma_{i,j}^{\textrm{D}}|\mathscr{I}}|^{2}\nonumber \\
 & \times\delta(E_{\mathscr{F}}-E_{\mathscr{I}}-T)\,,\label{eq:R^SD}
\end{align}
involves a statistical average of the initial states $\ket{\mathscr{I}}$
and a sum of all final states $\ket{\mathscr{F}}$ allowed by energy
conservation (imposed by the delta function). To incorporate relativistic
effects in atoms, the electron spin operator is expressed by the $4\times4$
Dirac spin matrices $\vec{\sigma}_{e}^{\textrm{D}}=2\vec{S}_{e}$,
with $\vec{\sigma}^{\textrm{D}}=\gamma^{0}\vec{\gamma}=\left(\begin{array}{cc}
\vec{\sigma} & 0\\
0 & \vec{\sigma}
\end{array}\right)$ where $\vec{\sigma}$ are the normal $2\times2$ Pauli matrices.
The transition operator contains all $Z$ electrons (the summation
of $i$), and spin operators of orthogonal $j$-directions contribute
incoherently (the summation of $j$). 

Our evaluation of $R_{\textrm{SD}}^{(\textrm{ion})}$ proceeds similarly
to our previous work on the SI case~\citep{Pandey:2018esq}. All
the essential formulae, including those new to the SD case, are given
in Appendix~\ref{sec:appendix} \textcolor{blue}{}%
. To emphasize once again the importance of relativistic and many-body
physics, we also highlight therein the differences from NR, independent-particle
approaches widely adopted in literature. 

\begin{figure*}
\begin{tabular}{cc}
\includegraphics[width=0.5\textwidth]{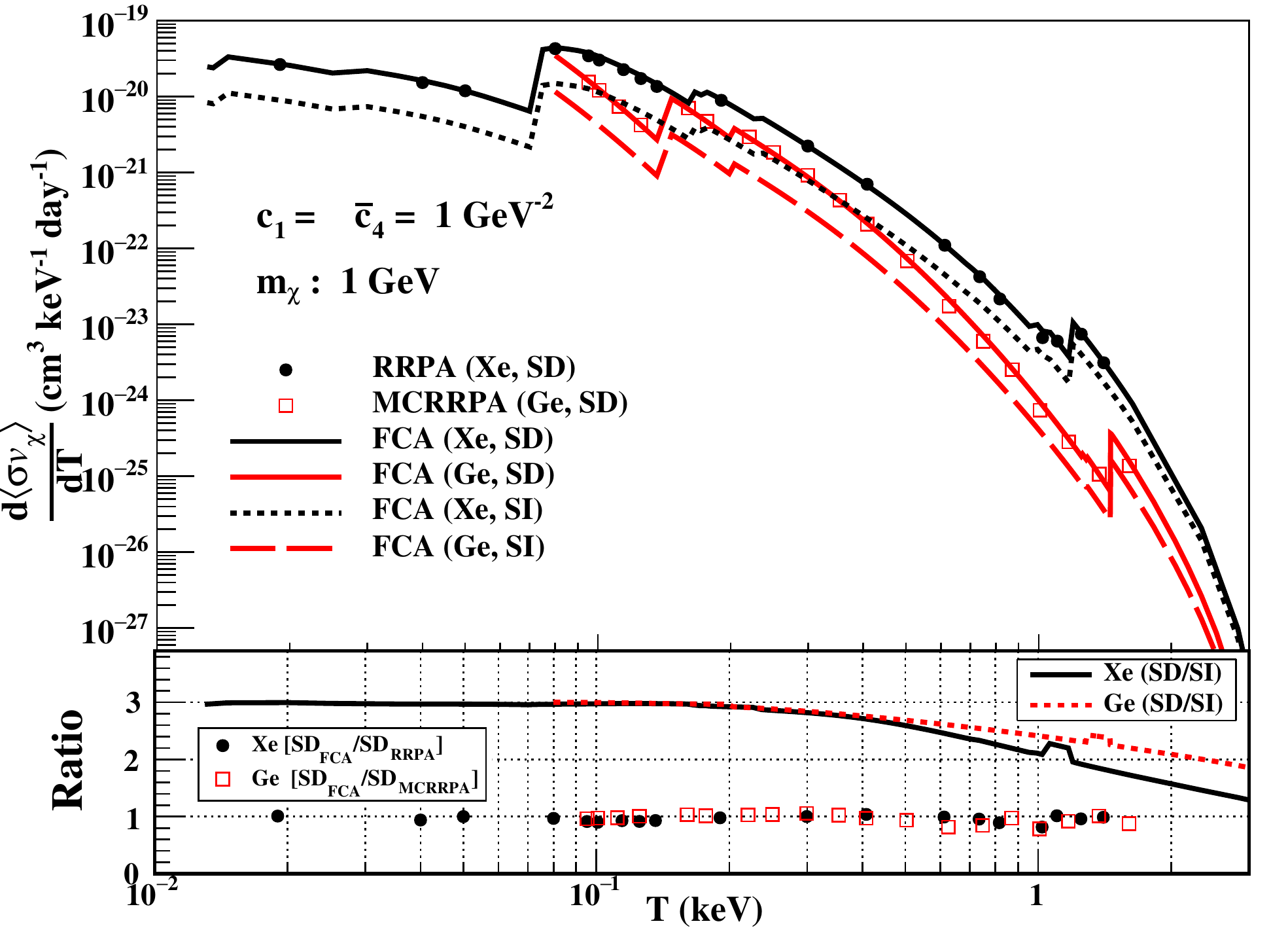} & \includegraphics[width=0.5\textwidth]{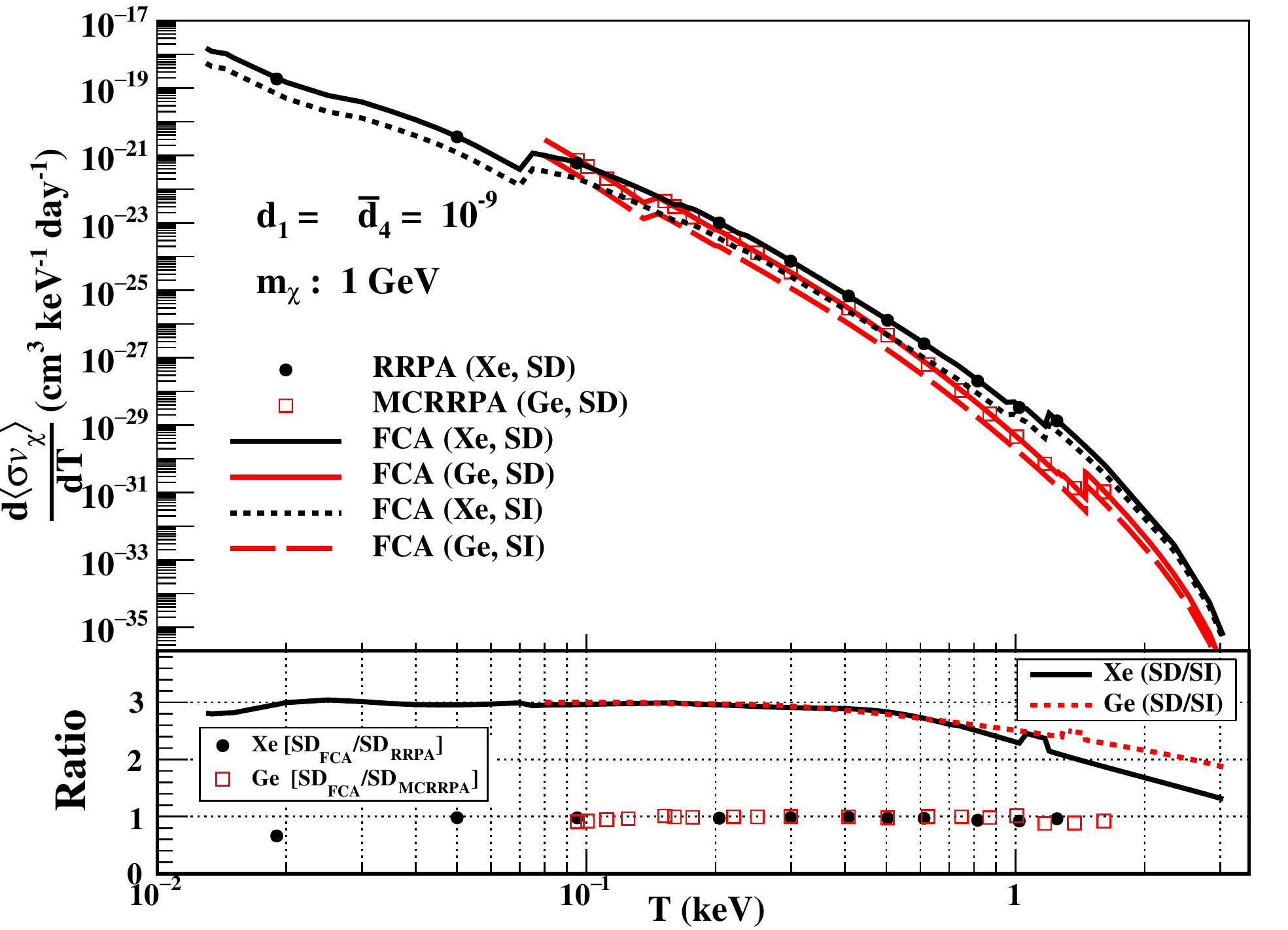}\tabularnewline
\end{tabular}\caption{Top: Averaged velocity-weighted differential cross sections $\frac{d\left\langle \sigma v_{\chi}\right\rangle }{dT}$
for ionizations of xenon (black) and germanium (red) by leading-order
spin-dependent (solid) and spin-independent (dashed) interactions
of short (left) and long (right) range, at $m_{\chi}=$1~GeV and
coupling constants of $c_{1}=\bar{c}_{4}=1/\textrm{GeV}^{-2}$ and
$d_{1}=\bar{d}_{4}=10^{-9}$. Points in circles (open squares) are
benchmark calculations by (MC)RRPA for xenon (germanium). Bottom:
Lines are ratios of $\frac{d\left\langle \sigma v_{\chi}\right\rangle }{dT}$
with the SD to the SI interactions. Points in closed circles (open
squares) are ratios of FCA to (MC)RRPA with the SD interaction for
xenon (germanium). }
\label{fig:dcs_ave}
\end{figure*}

While the computation of $R_{\textrm{SD}}^{(\textrm{ion})}$ involves
completely different multipole operators from its SI counterpart $R_{\textrm{SI}}^{(\textrm{ion})}$,
their definitions only differ formally in $R_{\textrm{SD}}^{(\textrm{ion})}$
having an additional spin operator $\vec{\sigma}^{\textrm{D}}$ and
the associated summation of three directions. Therefore, before presenting
our full numerical results, it is instructive to study the possible
relationship between them. For the purpose, we define a scaling factor
$\xi=R_{\textrm{SD}}^{(\textrm{ion})}/R_{\textrm{SI}}^{(\textrm{ion})}$.

First in Ref.~\citep{Kopp:2009et}, it was shown in DM scattering
off free electrons that $\xi^{(0)}=3$. Later in Ref.~\citep{Chen:2015pha},
the same conclusion was reached for hydrogen-like atoms that $\xi^{(\textrm{H})}=3$.
Recently, in Ref.~\citep{Catena:2019gfa}, the same conclusion was
generalized to many-electron atoms with two conditions: (i) the electron
basis wave functions are labeled by the principle (for bound states)
or energy (for continuum states), orbital, and magnetic quantum numbers:
$n$ or $k$ (with $E=k^{2}/2m_{e}$), $l$, $m_{l}$, and (ii) the
final states are purely one-particle-one-hole excitations from the
ground state. In such a NR, purely mean-field framework, it also yields
that $\xi^{(\textrm{NR})}=3$. Because a NR electron's orbital and
spin wave functions are factorized, the scaling factor of 3 can be
heuristic deducted from the multiplicity of the $\vec{\sigma}$ (for
the SD case) over the unity (for the SI case) operator, when the transition
involves only one independent electron. 

However, as there is spin-orbit interaction (SOI) in atoms, this scaling
relation can only be approximate at best. Its effects show up in all
parts of the response function. First, the SOI causes the energy shift
of two electron states of same $l$ but different total angular momenta
$j=l\pm1/2$, so the energy conserving delta functions are different.
Second, the electron basis wave functions (which are best solved by
relativistic mean field methods), their associated one-electron matrix
elements, and the density matrix (which contains the weights of all
atomic configurations that can be mixed by the residual interaction)
all have explicit dependence on $j$. Therefore, one chief goal of
this paper is to quantify to what extent that $\xi\approx3$ is a
good approximation to apply in DM-atom scattering. 

It should be pointed out that the analogous behavior for DM-nucleus
scattering is rather different. Unlike the atomic SOI, which is electromagnetic
with its strength characterized by the fine structure constant $\alpha_{\textrm{EM}}\approxeq1/137$,
the nuclear SOI is a component of the nuclear strong interaction,
and its strength is of the order of the strong coupling constant $\alpha_{\textrm{S}}\sim1$.
Consequently, there is no such simple scaling in DM-nucleus scattering
through the SD and SI DM-nucleon interactions (see, e.g., Ref.~\citep{Fitzpatrick:2012ix}). 

\section*{Results }

To get a prediction for the event rate at a detector with $N_{T}$
atoms 
\begin{equation}
\frac{dR}{dT}=\frac{\rho_{\chi}\,N_{T}}{m_{\chi}}\frac{d\braket{\sigma v_{\chi}}}{dT}\,,\label{eq:count-rate}
\end{equation}
the differential cross section is weighted and averaged by the standard
Maxwell-Boltzmann velocity distribution of DM~\citep{Lewin:1995rx},
$f(\vec{v}_{\chi})$, 
\begin{equation}
\frac{d\braket{\sigma v_{\chi}}}{dT}=\int_{v_{\min}}^{{v_{\textrm{max}}}}d^{3}v_{\chi}f(\vec{v}_{\chi})v_{\chi}\frac{d\sigma}{dT}\,,\label{eq:vchi_ave}
\end{equation}
with conventional choices of DM parameters (same as in Ref.~\citep{Pandey:2018esq}):
the local density $\rho_{\chi}=0.4$ GeV/cm$^{3}$, circular velocity
$v_{0}=220\,\textrm{km/s}$, escape velocity $v_{\textrm{esc}}=544\,\textrm{km/s}$,
averaged Earth relative velocity $v_{E}=232\,\textrm{km/s}$ so that
$v_{\textrm{max}}=v_{\textrm{esc}}+v_{E}$, and minimum $v_{\min}=\sqrt{2T/m_{\chi}}$.
We follow the same efficient velocity average scheme as in Refs.~\citep{Kopp:2009et,Essig:2011nj}
via an eta function $\eta(v_{\textrm{min}})=\left\langle v_{\chi}^{-1}\theta(v-v_{\textrm{min}})\right\rangle $
~\citep{Lewin:1995rx,Savage:2008er}. This simplification has been
explicitly verified in Ref.~\citep{Pandey:2018esq} for the same
kinematic region.

\begin{figure*}
\begin{tabular}{cc}
\includegraphics[width=0.5\textwidth]{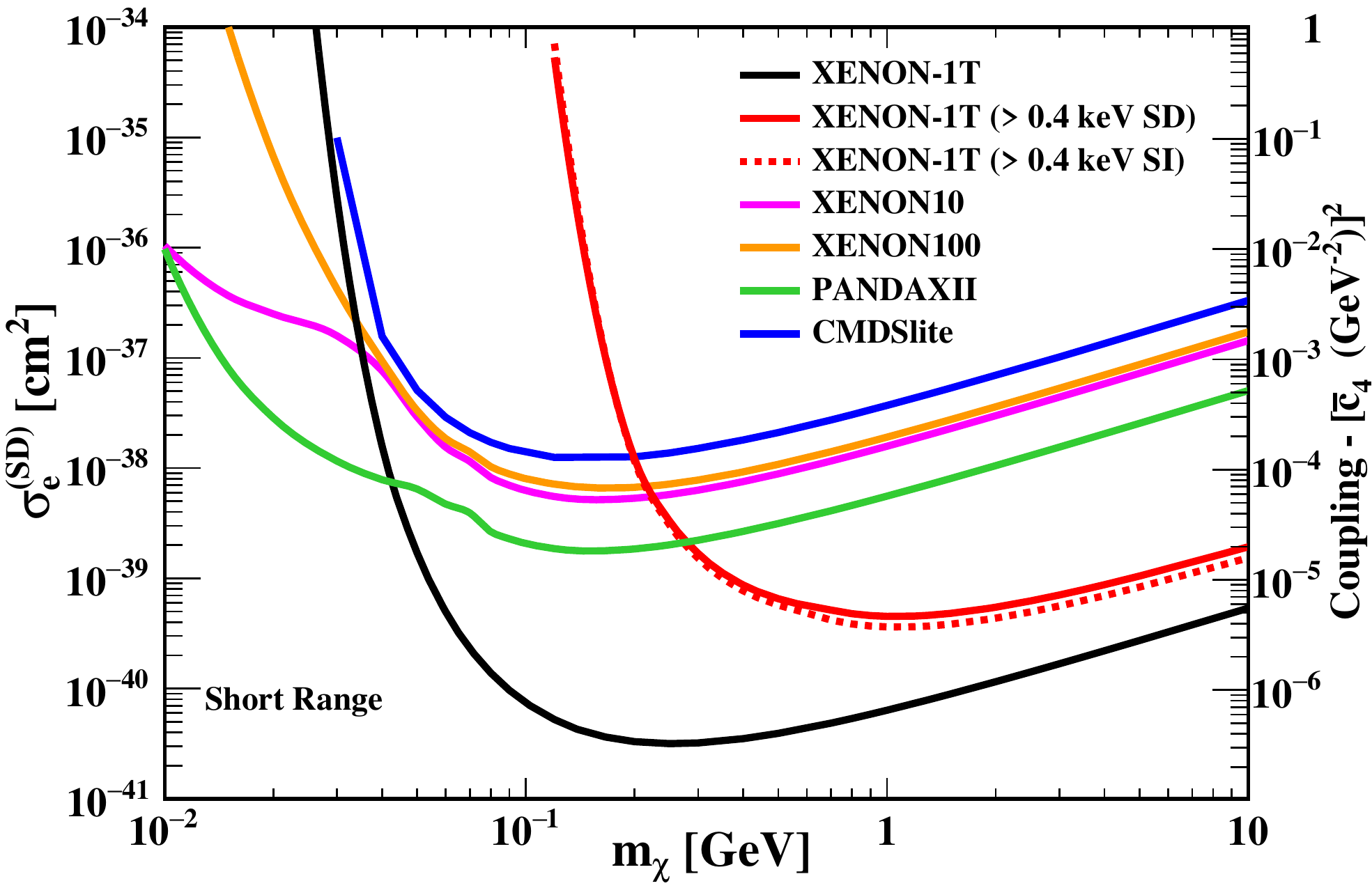} & \includegraphics[width=0.5\textwidth]{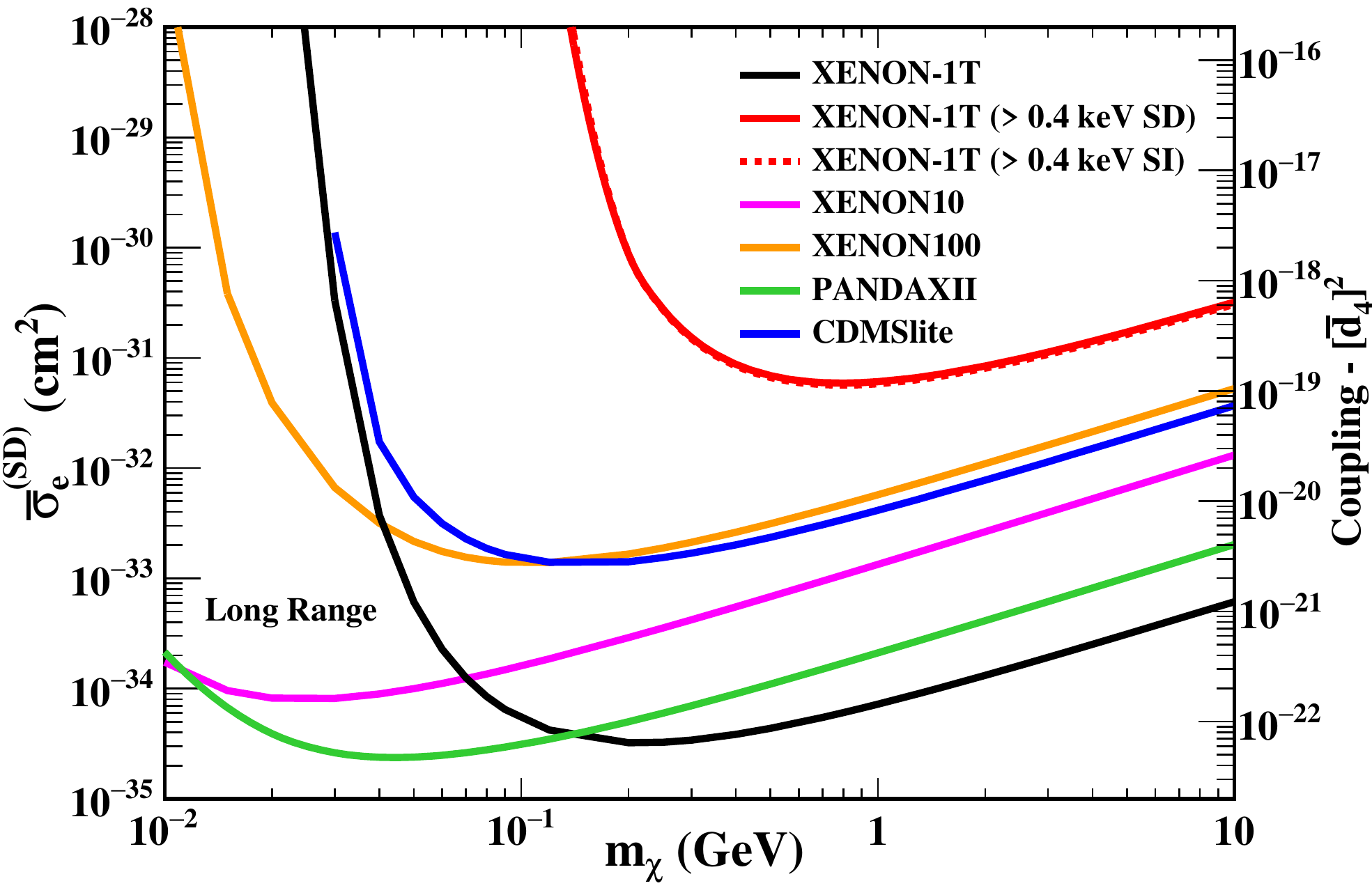}\tabularnewline
\end{tabular}\caption{Top: Exclusion limits at 90\% C.L. on the spin-dependent short- (left)
and long- (right) range DM-electron interactions as functions of $m_{\chi}$
derived from data of CDMSlite~\citep{Agnese:2017jvy}, XENON10~\citep{Essig:2012yx},
XENON100~\citep{Essig:2017kqs}, XENON1T~\citep{Aprile:2019xxb},
and PandaX-II~\citep{Cheng:2021fqb} data. The SD and SI limits of
XENON1T at 400 eV threshold are depicted as solid and dotted red lines,
respectively, to illustrate their expected recoil spectral shape are
different.~\label{fig:exclusion}}
\end{figure*}

In Fig.~\ref{fig:dcs_ave}, the results of $d\braket{\sigma v_{\chi}}/dT$
are shown for the cases with $m_{\chi}=1\,\textrm{GeV}$, $c_{1}=\bar{c}_{4}=1/\textrm{GeV}^{-2}$,
and $d_{1}=\bar{d}_{4}=10^{-9}$. The fully many-body calculations
through (multiconfiguration) relativistic random phase approximation,
(MC)RRPA, are performed at a few points (for them being computer-time-consuming)
labeled by circles (open squares) for xenon (germanium) atoms as benchmarks.
The main results in solid curves are obtained from a more efficient
approach: the frozen core approximation (FCA), which is in the spirit
of an independent particle approximation but with a more realistic
mean field to compute the ionized electron wave function (details
in Ref.~\citep{Pandey:2018esq}). In general, FCA shows good agreement
with (MC)RRPA benchmark points except near edge energies. Considering
the $5\%$ accuracy of (MC)RRPA achieved in photoabsorption cross
sections with $T\ge$12 and 80~eV for xenon~\citep{Chen:2016eab}
and germanium~\citep{Chen:2013lba,Chen:2014ypv}, respectively, we
conservatively assign an overall theory error of $20\%$. (The theory
error can be reduced to 5\%, if we perform (MC)RRPA calculations throughout
the work without the limitation of computer resources.) Complete data
of $d\braket{\sigma v_{\chi}}/dT$ with different $m_{\chi}$ are
provided in the folder of ancillary files.

Data from CDMSlite~\citep{Agnese:2017jvy}, XENON10~\citep{Essig:2012yx},
XENON100~\citep{Essig:2017kqs}, XENON1T~\citep{Aprile:2019xxb},
and PandaX-II~\citep{Cheng:2021fqb} are analyzed against the theory
predictions of $dR/dT$ in the same way as in Ref.~\citep{Pandey:2018esq}).
In Fig.~\ref{fig:exclusion}, the exclusion limits on the coupling
constants $\bar{c}_{4}$ and $\bar{d}_{4}$, and equivalently the
SD DM-electron total cross sections $\sigma_{e}^{(\textrm{SD})}=3\bar{c}_{4}^{2}\mu_{\chi e}^{2}/\pi$
and $\bar{\sigma}_{e}^{(\textrm{SD})}=3\bar{d}_{4}^{2}\mu_{\chi e}^{2}/\left(\pi(m_{e}\alpha)^{4}\right)$,
where $\mu_{\chi e}=m_{\chi}m_{e}/(m_{\chi}+m_{e})$, are presented
on the right and left $y$-axis, respectively. For $m_{\chi}$ larger
than 42 and 140 MeV, XENON1T provides the most stringent limits for
the short- and long-range interactions, respectively, with its ton-year
scale exposure. Particularly in the range between $0.1-10\,\textrm{GeV}$,
the limits on $\sigma_{e}^{(\textrm{SD})}<10^{-41}-10^{-40}\,\textrm{cm}^{2}$
are of similar order of magnitude as the ones of $\sigma_{n,p}^{(\textrm{SD})}$
at slightly bigger $m_{\chi}$. This is an important complement to
mainstream searches of weakly-interacting massive particles through
DM-nucleus scattering: not only because of some overlapping in mass
range (1-10 GeV), but also for classes of DM which can be both hadro-
and lepto-phillic. For smaller $m_{\chi}$ down to 10 MeV, PandaX-II
instead set better limits because of its lower ER threshold at $80\,\textrm{eV}$.
It defines the exclusion boundary at low $m_{\chi}$ for having lower
background $\sim2.64\times10^{-4}\,\textrm{cpd/kg/keV}$ compared
to $\sim0.2$ of XENON10, despite the latter has an even lower threshold
at $13.8\,\textrm{eV}$. 

We emphasize here the derived exclusion limits depend critically on
the theory predictions for the DM scattering event rates. Substantial
differences exist between our relativistic many-body approach and
several NR mean-field approaches Refs.~\citep{Essig:2011nj,Essig:2012yx,Essig:2017kqs,Agnes:2018oej,Catena:2019gfa}
including the package QEdark that was used in the data analyses of
Refs.~\citep{Aprile:2019xxb,Cheng:2021fqb}. Detailed comparisons,
reasons that cause the differences, and justifications of our approach
are presented in Ref.~\citep{Pandey:2018esq}. Based on similar arguments,
one would generically anticipate weaker limits on the short-range
interaction in the high-mass region, and tighter limits on the long-range
interaction in the low-mass region, when above-mentioned approaches
are applied. The main reasons are: (i) the relativistic effects substantially
increase the scattering event rate at high $T$, and (ii) the many-body
effects play an important role at all distances, including the Coulomb
screening at long distances which substantially decrease the scattering
rate at very low $T$. 

Not included in our data analyses are several low-threshold semiconductor
experiments, which already set limits on $\sigma_{e}^{(\textrm{SI})}$
below 10~MeV~\citep{Agnese:2018col,Crisler:2018gci,Abramoff:2019dfb,Barak:2020fql,Aguilar-Arevalo:2019wdi,Arnaud:2020svb}
and have the potential to compete with liquid noble-gas detectors
above the 10~MeV region, particularly for the long-range interactions,
in the future. The associated many-body physics of condensed matter
is beyond the scope of this work, but has been taken in Refs.~\citep{Graham:2012su,Essig:2015cda,Hochberg:2021pkt,Knapen:2021run,Knapen:2021bwg,Catena:2021qsr,Griffin:2021znd}
for the SI interaction, and recently in Refs.~\citep{Trickle:2019ovy,Trickle:2020oki,Catena:2021qsr,Griffin:2021znd}
that also include the SD interactions. Currently, there are discrepancies
among predictions from different theoretical schemes.

\section*{Disentanglement of SD and SI interactions}

Conventional DM detectors only measure recoil energy. Detector responses
to SI and SD signatures must be different in order to disentangle
them. Therefore, if the NR scaling $\xi^{(\textrm{NR})}=3$ is valid,
the SD and SI signals are indistinguishable in the measurable recoil
energy spectrum shapes and the exclusion limits of $\sigma_{e}^{(\textrm{SD})}$
and $\sigma_{e}^{(\textrm{SI})}$ would have identical $m_{\chi}$
dependence. 

In the bottom panel of Fig.~\ref{fig:dcs_ave}, this statement is
examined through the $\bar{\xi}$ parameter, which is the ratio of
$\frac{d\left\langle \sigma v_{\chi}\right\rangle }{dT}$ with the
SD to the SI interaction, for the case of $m_{\chi}=1\,\textrm{GeV}$.
The overbar in $\bar{\xi}$ indicates the responses functions are
integrated over $q$ and averaged over the DM flux in $\frac{d\left\langle \sigma v_{\chi}\right\rangle }{dT}$.
At $T\lesssim200\,\textrm{eV}$, the scaling relation $\bar{\xi}\approx3$
works well for both xenon and germanium. The scaling deviation starts
to grow as $T$ increases, and the larger deviations in xenon than
germanium demonstrates the effects of a stronger SOI in an atom of
higher $Z$. We note that the SOI is only part of the relativistic
corrections in atoms. Therefore, even though it has been shown that
relativistic corrections become sizable for $T\gtrsim200\,\textrm{eV}$
in DM-xenon scattering~\citep{Pandey:2018esq}, the departure of
$\bar{\xi}$ from the NR prediction $\bar{\xi}^{\textrm{(NR)}}=3$
is comparatively modest.

It follows that precision spectral shape measurements at high energy
allow SD and SI to be distinguished in the case of positive discovery
signals. As SD limits are driven by detector thresholds lower than
200~eV, the exclusion curves of Fig.~\ref{fig:exclusion} differ
only slightly from the ones for SI given in Fig.~2 of Ref.~\citep{Pandey:2018esq}.
If a higher cutoff is applied, as the example shown in Fig.~\ref{fig:exclusion}
with 400-eV cutoff to XENON1T data, noticeable difference shows up.

There are intriguing possibilities of experimental signatures unique
to SD interactions without SI background. For example, in a spin polarizable
target and with known spin states of the ionized electrons, only part
of the SD interaction $\sum_{\lambda=\pm1}(-1)^{\lambda}S_{\chi,\lambda}S_{e,-\lambda}$
(where $\lambda$ denotes the spherical component) can generate signals
as spin-flips. There are intense recent interests to seek solutions
in condensed phases of matter which have natural spin, or magnetic,
order. For instance, a ferrimagnetic material: yttrium iron garnet
has been proposed as a candidate target~\citep{Trickle:2019ovy,Trickle:2020oki}
for light DM searches with $m_{\chi}\le10\,\textrm{MeV}$. The SD
DM-electron interaction can perturb this spin system and generate
magnons, which are the low-energy quanta of spin waves and experimentally
measurable. 

Motivated by the role of SOI, another approach is to realize experimental
configurations where spins couple strongly to orbital angular momenta.
For atoms, the fine structure constant is fixed, so that SOI enhancement
can only be found in the inner-shell electrons, which have higher
ionization energies. The novel Dirac materials do have the desired
property that the effective fine structure constants can be larger
than $\alpha_{\textrm{EM}}$. Take graphene as an example: a naive
dimensional analysis yields $\alpha_{g}=\alpha_{\textrm{EM}}/v_{F}\approx2.2$,
adopting a generic Fermi velocity at Dirac cone $v_{F}\approx10^{6}\,\textrm{m/s}$.
While recent calculations~\citep{Siegel:2011gra} and measurements~\citep{Reed:2010gra,Gan:2016gra}
reported smaller values in between 0.1 and 1 (mostly due to the suppression
by Coulomb screening), they are still significantly bigger than $\alpha_{\textrm{EM}}$.
The smallness of $v_{F}$ compared to the speed of light is the most
important factor that lowers the bar to reach the relativistic limit.
Unfortunately, the intrinsic SOI in graphene is found to be small~\citep{CastroNeto:2009zz},
so whether it could generate a sufficiently different SD response
requires more study. More promising candidates are topological insulators~\citep{Hasan:2010xy,Qi:2011zya}
and transition-metal dichalcogenides~\citep{Zhu:2011so}, whose structures
critically depend on SOIs. In addition, among a few Dirac materials
being proposed to search for SI DM-electron interactions, the semimetal
$\textrm{ZrTe}_{5}$~\citep{Hochberg:2017wce,Geilhufe:2019ndy,Coskuner:2019odd}
and organic crystal $(\textrm{BEDT-TTF})\cdot\textrm{Br}$~\citep{Geilhufe:2018gry}
show corrections from SOIs in band structure calculations. 

\section*{Summary}

The SD DM-electron interactions is an important, but less-attended
subject in direct DM searches. Its studies complement the very active
research on the SI interactions, and together they can provide a more
comprehensive understanding about the nature of DM and its interactions
with matter. We derive in this paper, for the first time, limits on
the SD DM-electron cross sections at leading order with state-of-the-arts
atomic many-body calculations and current best experiment data. As
a result of spin-orbit interaction, the spectral shapes of SD and
SI recoil spectra differ at high energies, which provides a way of
differentiating them experimentally. In addition, signals unique to
SD interactions and systems of enhanced SOIs that can be found in
novel phases of matter may offer more direct experimental probes for
SD from sub-GeV DM searches in the future.
\begin{acknowledgments}
This work was supported in part under Contract Nos.~106-2923-M-001-006-MY5,
108-2112-002-003-MY3, 108-2112-M-259-003, and 109-2112-M-259-001 from
the Ministry of Science and Technology, 2019-20/ECP-2 and 2021/TG2.1
from the National Center for Theoretical Sciences, and the Kenda Foundation
(J. W. C.) of Taiwan; and the Canada First Research Excellence Fund
through the Arthur B. McDonald Canadian Astroparticle Physics Research
Institute (C. P. W.).
\end{acknowledgments}

\bibliography{DM-e_SD_arXiv}

\newpage

\appendix

\section{Atomic Many-body Response Functions~\label{sec:appendix}}

It is a well-acknowledged fact that a detector's response to scattering
of nonrelativistic sub-GeV DM involves many-body physics at atomic,
molecular, or condensed matter scale, depending on the configuration
of a detector. In this Appendix, we summarize the important steps
and formulae in a truly many-body calculation of the atomic response
functions, and illustrate the differences to approaches based solely
on independent particle pictures, either relativistically or non-relativistically.
We shall follow the notations and terminology of Ref.~\citep{Fitzpatrick:2012ix}
(a seminal work on the nuclear responses to DM-nucleus scattering)
as much as possible. However, a necessary addition is using Dirac
spinors for electrons, as relativistic effects are important and manifest
in atomic physics. A further note is while a formulation in molecular
or condensed matter systems proceeds similarly in spirit, there are
substantial changes of basic elements and notations which we leave
for future work. 

For brevity, consider only the short-range DM-electron interactions
\begin{equation}
\mathscr{L}^{(\textrm{LO})}=c_{1}\left(\chi^{\dagger}\mathbbm{1}_{\chi}\chi\right)\cdot\left(e^{\dagger}\mathbbm{1}_{e}\right)+c_{4}\left(\chi^{\dagger}\vec{S}_{\chi}\chi\right)\cdot\left(e^{\dagger}\vec{S}_{e}e\right)\,,
\end{equation}
with both leading-order SI ($c_{1}$) and SD ($c_{4}$) terms included.
The unpolarized differential scattering cross section in the laboratory
frame is given by 
\begin{equation}
\frac{d\sigma}{dT}^{(\textrm{ion})}=\frac{1}{2\pi v_{\chi}^{2}}\int qdq\,c_{1}^{2}R^{(\textrm{SI})}(T,q)+\bar{c}_{4}^{2}R^{(\textrm{SD})}(T,q)\,,\label{eq:dcs_LO}
\end{equation}
where $\bar{c}_{4}=\sqrt{s_{\chi}(s_{\chi}+1)/12}\,c_4$ that absorbs the
factors due to DM spin, $s_{\chi}$, and electron spin $s_{e}=1/2$.
The SI and SD detector responses are encoded in two functions

\begin{subequations}
\begin{align}
R_{\textrm{SI}}^{(\textrm{ion})}(T,q)= & \sum_{\mathscr{F}}\overline{\sum_{\mathscr{I}}}\,|\braket{\mathscr{F}|\sum_{i=1}^{Z}e^{i\vec{q}.\vec{r}_{i}}|\mathscr{I}}|^{2}\,\delta(E\ldots)\,,\\
R_{\textrm{SD}}^{(\text{ion})}(T,q)= & \sum_{\mathscr{F}}\overline{\sum_{\mathscr{I}}}\sum_{k}\,|\braket{\mathscr{F}|\sum_{i=1}^{Z}e^{i\vec{q}.\vec{r}}\sigma_{i,k}^{\mathrm{D}}|\mathscr{I}}|^{2}\,\delta(E\ldots)\,,
\end{align}
\end{subequations}
where the delta function imposes energy conservation $E_{\mathscr{F}}=E_{\mathscr{I}}+T$.

A substantial simplification to the problem can usually be achieved
by multipole expansion. First, because total angular momentum and
its $z$-projection, are good quantum numbers, so we label the initial
and final states as $\ket{\mathscr{I}}=\ket{I,J_{I}M_{J_{I}}}$ and
$\ket{\mathscr{F}}=\ket{F,J_{F}M_{J_{F}}}$, respectively, where $I$
and $F$ are collective labels for other quantum numbers. Second,
each multipole operator has its definite spherical rank, $J$, component,
$M_{J}$, and parity, so the selection rules help to organize all
allowed transitions in a systematic way. Third, by the Wigner-Eckart
theorem, the summation over $M_{J_{F}}$, $M_{J_{I}}$, and $M_{J}$
can be carried out easily, and what remain to be calculated are the
reduced matrix elements. Most important of all, for low energy processes,
the expansion in spherical rank converges efficiently. 

Following the notation of Ref.~\citep{Fitzpatrick:2012ix}, the SI
and SD response functions can now be decomposed as 

\begin{subequations}
\begin{align}
R_{\textrm{SI}}^{(\textrm{ion})}(T,q)= & \sum_{FJ_{F}}\overline{\sum_{IJ_{I}}}\frac{4\pi}{2J_{I}+1}\sum_{J=0}\left|\left\langle F,J_{F}\left\Vert \hat{M}_{J}(q)\right\Vert I,J_{I}\right\rangle \right|^{2}\delta(E\ldots)\,,\\
R_{\textrm{SD}}^{(\textrm{ion})}(T,q)= & \sum_{FJ_{F}}\overline{\sum_{IJ_{I}}}\frac{4\pi}{2J_{I}+1}\left\{ \sum_{J=1}\left|\left\langle F,J_{F}\left\Vert \hat{\Sigma}_{J}(q)\right\Vert I,J_{I}\right\rangle \right|^{2}\right.\nonumber \\
 & \left.+\sum_{J=1}\left|\left\langle F,J_{F}\left\Vert \hat{\Sigma}_{J}^{'}(q)\right\Vert I,J_{I}\right\rangle \right|^{2}+\sum_{J=0}\left|\left\langle F,J_{F}\left\Vert \hat{\Sigma}_{J}^{''}(q)\right\Vert I,J_{I}\right\rangle \right|^{2}\right\} \delta(E\ldots)\,,
\end{align}
\end{subequations}
and the relevant four types of multipole operators are 

\begin{subequations}
\begin{align}
\hat{M}_{J}^{M_{J}}(q) & =\sum_{i=1}^{Z}j_{J}(qr_{i})Y_{J}^{M_{J}}(\Omega_{r_{i}})\,,\\
\hat{\Sigma}_{J}^{M_{J}}(q) & =\sum_{i=1}^{Z}j_{J}(qr_{i})\vec{Y}_{JJ}^{M_{J}}(\Omega_{r_{i}})\cdot\vec{\sigma}_{i}^{\textrm{D}}\,,\\
\hat{\Sigma}_{J}^{'M_{J}}(q) & =\sum_{i=1}^{Z}\left\{ -\sqrt{\frac{J}{2J+1}}j_{J+1}(qr_{i})\vec{Y}_{JJ+1}^{M_{J}}(\Omega_{r_{i}})+\sqrt{\frac{J+1}{2J+1}}j_{J-1}(qr_{i})\vec{Y}_{JJ-1}^{M_{J}}(\Omega_{r_{i}})\right\} \cdot\vec{\sigma}_{i}^{\textrm{D}}\,,\\
\hat{\Sigma}_{J}^{''M_{J}}(q) & =\sum_{i=1}^{Z}\left\{ \sqrt{\frac{J+1}{2J+1}}j_{J+1}(qr_{i})\vec{Y}_{JJ+1}^{M_{J}}(\Omega_{r_{i}})+\sqrt{\frac{J}{2J+1}}j_{J-1}(qr_{i})\vec{Y}_{JJ-1}^{M_{J}}(\Omega_{r_{i}})\right\} \cdot\vec{\sigma}_{i}^{\textrm{D}}\,,
\end{align}
\end{subequations}
where $j_{J}(qr)$ is the spherical Bessel function, $Y_{J}^{M_{J}}(\Omega_{r})$
the spherical harmonics of solid angle $\Omega_{r}$, and $\vec{Y}_{JL}^{M_{J}}(\Omega_{r})$
the vector spherical harmonics formed by recoupling of $Y_{L}^{M_{L}}(\Omega_{r})$
and the unit vector $\hat{r}$, whose spherical projection is proportional
to $Y_{1}^{\lambda}$.

A general feature of all these matrix elements to be computed is they
involve many-body states $\ket{\mathscr{I}}$ and $\ket{\mathscr{F}}$,
but the operators are one-body, i.e, only one electron makes transition
at one time. Such matrix elements of a generic one-body operator $\hat{O}^{(1)}=\sum_{i}\hat{o}^{(i)}$
can be compactly written by second quantization that 
\begin{equation}
\braket{\mathscr{F}|\hat{O}^{(1)}|\mathscr{I}}=\sum_{\alpha\beta}\Psi_{\mathscr{FI}}(\alpha,\beta)\braket{\alpha|\hat{o}|\beta}\,,
\end{equation}
where the one-body density matrix 
\begin{equation}
\Psi_{\mathscr{FI}}(\alpha,\beta)=\braket{\mathscr{F}|\hat{c}_{\alpha}^{\dagger}\hat{c}_{\beta}|\mathscr{I}}
\end{equation}
gives the statistical weight of annihilating an electron with a quantum
state labeled by $\beta$ and creating an electron with a quantum
state labeled by $\alpha$ in this transition, and the total transition
amplitude is the weighted sum of each single-particle transition with
an amplitude given by $\braket{\alpha|\hat{o}|\beta}$ (the reference
to the $i$th electron is dropped because electrons are identical.

Implementing the above scheme to a spherical basis with spherical
operators is standard but too lengthy to reproduce here, but we should
mention two basic ingredients. First, all single-particle states have
good quantum labels of total angular momentum, so one can label them
as $\ket{\alpha}=\ket{aj_{a}m_{j_{a}}}$ and $\ket{\beta}=\ket{bj_{b}m_{j_{b}}}$.
Second, by recoupling the creation and annihilation operator $\hat{c}_{\alpha}^{\dagger}$
and $\hat{c}_{\beta}$, a spherical tensor operator of rank $J$ and
component $M_{J}$ can be formed 
\begin{equation}
[\hat{c}_{\alpha}^{\dagger}\otimes\hat{c}_{\beta}]_{J}^{M_{J}}=\sum_{m_{j_{a}}m_{j_{b}}}\braket{j_{a}j_{b};JM_{J}|j_{a}j_{b};m_{j_{a}}m_{_{j_{b}}}}\hat{c}_{\alpha}^{\dagger}\hat{c}_{\beta}\,,
\end{equation}
where the braket is the standard Clebsch-Gordan coefficient. Consequently,
the reduced matrix element of a spherical operator can be recast as
\begin{equation}
\braket{F,J_{F}||\hat{O}_{J}^{(1)}||I,J_{I}}=\frac{1}{\sqrt{2J+1}}\sum_{aj_{a}bj_{b}}\bar{\Psi}_{FJ_{F};IJ_{I}}^{J}(aj_{a},bj_{b})\braket{aj_{a}||\hat{o}_{J}||bj_{b}}\,,
\end{equation}
and the resulting density matrix is also reduced 
\begin{equation}
\bar{\Psi}_{FJ_{F},IJ_{I}}^{J}(aj_{a},bj_{b})=\left\langle F,J_{F}\left\Vert [\hat{c}_{\alpha}^{\dagger}\otimes\hat{c}_{\beta}]_{J}\right\Vert I,J_{I}\right\rangle \,,
\end{equation}
in the sense that it no longer depends on total angular momentum substates
$m_{j_{a}}$ and $m_{j_{b}}$, and the spherical multipole component
$M_{J}$. 

At this point we should note that as long as many-body wave functions
can be obtained exactly, the choice of a complete set of single-particle
basis states $\{\ket{\alpha},\ket{\beta},\ldots\}$ is irrelevant,
because the density matrix is also known exactly. However, this does
not happen in practical cases, and the arts of many-body methods is
mostly about choosing good sets of basis states and solving the density
matrix reliably. The approach by Ref.~\citep{Fitzpatrick:2012ix},
the nuclear shell model, is built upon a simple basis spanned by the
harmonic oscillator eigenstates, which have nice properties that yield
algebraic results of single-particle matrix elements. This simplistic
mean field is compensated by diagonalizing the residual interaction
in a large, but still truncated, model space with proper care in using
the effective operators. Our approach, the (multiconfiguration) relativistic
random phase approximation ((MC)RRPA), proceeds in a rather different
direction. The mean field is first solved self-consistently by the
Dirac-Hartree-Fock equation, so the single-particle basis states are
optimized, i.e, they already include a large part of many-body physics.
Then the effects due to the residual interactions are accounted for
by the random phase approximation. The resulting density matrices,
though not as dense as the ones of shell model approaches, are certainly
not sparse when residual correlation induces substantial configuration
mixing.

\subsection{Independent Particle Approximation}

Full-scale many-body calculations of the atomic responses by (MC)RRPA,
or other methods that take residual interactions into account, usually
have high demand of computing resources. Therefore, it is desirable
to seek approximations that perform reasonably good but more efficiently.
For the purpose, the independent particle approximation (IPA) based
on mean field approaches serves as a good starting point. 

In such a picture, the ground state of an atom is approximated by
filling all electrons to lowest $Z$ single-particle orbitals, which
are solutions of the mean field equation, and the resulting many-body
wave function, as a Slater determinant (or a linear combination of
several equivalent configurations for open-shell atoms), can successfully
describe a lot, if not all, static ground state properties of an atom.
When considering the action of a one-body operator on this assembly
of $Z$ noninteracting electrons, called core electrons, then the
excited state is simply one electron leaving the core and occupying
a higher energy level (for ionization, a positive-energy state in
continuum). Conventionally, this final state is called a one-particle-one-hole
(1p1h) excitation of the initial ground state, which is taken to be
the physical vacuum $\ket{0}$ of zero particle and zero hole, and
also zero total angular momentum $J_{I}=0$. A final state of $J_{F}=J$
that corresponds to one bound electron in the orbital $\ket{n_{i}l_{i}j_{i}}$
being ionized to a continuum orbital $\ket{k_{f}l_{f}j_{f}})$ can
be completely specified as $\ket{(k_{f}l_{f}j_{f})(n_{i}l_{i}j_{i})^{-1},J_{F}=J}=\ket{\textrm{1p1h};J}$.
As a result, the density matrix in the independent particle picture
is extremely simple: 
\begin{equation}
\left\langle \textrm{1p1h};J\left\Vert [\hat{c}_{\alpha}^{\dagger}\otimes\hat{c}_{\beta}]_{J}\right\Vert 0;0\right\rangle =\sqrt{2J+1}\delta_{a,k_{f}l_{f}}\delta_{j_{a}j_{f}}\delta_{b,n_{i}l_{i}}\delta_{j_{b}j_{i}}\,,
\end{equation}
and the many-body matrix element is reduced to the one with single-particle
states 
\begin{equation}
\braket{\textrm{1p1h};J||\hat{O}_{J}^{(1)}||0;0}=\braket{k_{f}l_{f}j_{f}||\hat{o}_{J}||n_{i}l_{i}j_{i}}\,,
\end{equation}
where $|j_{f}-j_{i}|\le J\le j_{f}+j_{i}$.

The spin-angular parts of the relevant single-particle matrix elements
can be worked out algebraically as follows. First, we express the
relativistic single-particle orbital functions in coordinate space
as 
\begin{equation}
\phi_{\varepsilon ljm}(\vec{r})=\frac{1}{r}\left(\begin{array}{c}
g_{\varepsilon lj}(r)\mathscr{Y}_{jl}^{m}(\Omega_{r})\\
-if_{\varepsilon lj}(r)(\vec{\sigma}\cdot\hat{r})\mathscr{Y}_{jl}^{m}(\Omega_{r})
\end{array}\right)\,,\label{eq:rel_wf}
\end{equation}
where $\varepsilon$ refers to $n\,(k)$ for the bound (free) electron,
and $\mathscr{Y}_{jl}^{m}(\Omega_{r})$ the spin-angular function.
Then the reduced matrix elements are simplified by the following two
formulae: 

\begin{subequations}
\begin{align}
\braket{\varepsilon_{f}l_{f}j_{f}||j_{J}(qr)Y_{J}(\Omega_{r})||\varepsilon_{i}l_{i}j_{i}} & =\frac{(-1)^{J+j_{i}+1/2}}{\sqrt{4\pi}}[l_{f}][l_{i}][j_{f}][j_{i}][J]\left\{ \begin{array}{ccc}
l_{f} & j_{f} & \nicefrac{1}{2}\\
j_{i} & l_{i} & J
\end{array}\right\} \left(\begin{array}{ccc}
l_{f} & J & l_{i}\\
0 & 0 & 0
\end{array}\right)\times\left\langle \varepsilon_{f}l_{f}j_{f}\left|j_{J}(qr)\right|\varepsilon_{i}l_{i}j_{i}\right\rangle \\
\braket{\varepsilon_{f}l_{f}j_{f}||j_{L}(qr)\vec{Y}_{JL}(\Omega_{r})\cdot\vec{\sigma}^{\textrm{D}}||\varepsilon_{i}l_{i}j_{i}} & =\frac{(-1)^{l_{f}}\sqrt{6}}{\sqrt{4\pi}}[l_{f}][l_{i}][j_{f}][j_{i}][L][J]\left\{ \begin{array}{ccc}
l_{f} & l_{i} & L\\
\nicefrac{1}{2} & \nicefrac{1}{2} & 1\\
j_{f} & j_{i} & J
\end{array}\right\} \left(\begin{array}{ccc}
l_{f} & L & l_{i}\\
0 & 0 & 0
\end{array}\right)\times\left\langle \varepsilon_{f}l_{f}j_{f}\left|j_{L}(qr)\right|\varepsilon_{i}l_{i}j_{i}\right\rangle 
\end{align}
\end{subequations}
where the shorthand notation $[x]=\sqrt{2x+1}$, and the standard
forms of the Wigner 3-$j$, 6-$j$, and 9-$j$ symbols are used. 

The most important structure information is contained in the radial
integrals 
\begin{equation}
\left\langle \varepsilon_{f}l_{f}j_{f}\left|j_{J}(qr)\right|\varepsilon_{i}l_{i}j_{i}\right\rangle =\int dr\,j_{J}(qr)\,(g_{\varepsilon_{f}l_{f}j_{f}}(r)g_{\varepsilon_{i}l_{i}j_{i}}(r)+f_{\varepsilon_{f}l_{f}j_{f}}(r)f_{\varepsilon_{i}l_{i}j_{i}}(r))\,,
\end{equation}
which depend on the single-particle orbital wave functions. For the
electrons in the core, as the mean field methods based on variational
principles are designed to get a best approximation for the ground
state energy, the resulting wave functions can be taken with certain
confidence. However, it is more problematic for the ionized electron,
and such a highly excited state goes beyond the reach of conventional
mean field design. There are many methods applied to DM-atom scattering,
including plane wave, hydrogen-like, and frozen core approximations.
We refer readers to Ref.~\citep{Pandey:2018esq} for more discussions
and comparison of these methods, and only point out here that even
though our main results in this paper are obtained with a frozen core
approximation (FCA), the latter approach is justified by a few benchmark
calculations using (MC)RRPA.

\subsection{Nonrelativistic Limit}

By further taking the nonrelativistic (NR) limit to the above independent
particle approximation, we now show explicitly how our formulation
can be reduced to the conventional form widely used in the literature,
and recover the NR scaling factor $\xi^{(\textrm{NR})}=3$. 

In numerical computations, the NR limit can be implemented by taking
the speed of light to infinity. Theoretically, the most important
features are (i) the Dirac spinor now has a vanishing small component
and collapses to a Pauli spinor, i.e., all functions $f_{\varepsilon lj}(r)$
can be taken to zero, and (ii) the radial wave functions now only
depends on $\varepsilon$ and $l$, since there is no longer spin-orbit
interaction to break the $j=l\pm1/2$ degeneracy. As a result, the
only change in our formulation is in the radial integral 
\begin{equation}
\left\langle \varepsilon_{f}l_{f}j_{f}\left|j_{J}(qr)\right|\varepsilon_{i}l_{i}j_{i}\right\rangle \rightarrow\left\langle \varepsilon_{f}l_{f}\left|j_{J}(qr)\right|\varepsilon_{i}l_{i}\right\rangle _{(\textrm{NR})}=\int dr\,j_{J}(qr)\,u_{\varepsilon_{f}l_{f}}(r)u_{\varepsilon_{i}l_{i}}(r)\,,
\end{equation}
where $u_{\varepsilon lj}(r)$'s are the NR radial wave functions.

In the NR-IPA scheme, the SI response function 
\begin{align}
\mathcal{R}{}_{\textrm{SI}}^{(\textrm{ion})}(T,q) & =\sum_{k_{f}l_{f}j_{f}}\sum_{n_{i}l_{i}j_{i}}\sum_{L=0}4\pi\left|\braket{k_{f}l_{f}j_{f}||j_{L}(qr)Y_{L}(\Omega_{r})||n_{i}l_{i}j_{i}}\right|^{2}\delta(E\ldots)\,,\nonumber \\
 & =\sum_{k_{f}l_{f}}\sum_{n_{i}l_{i}}\sum_{L=0}2[l_{f}]^{2}[l_{i}]^{2}[L]^{2}\left(\begin{array}{ccc}
l_{f} & L & l_{i}\\
0 & 0 & 0
\end{array}\right)^{2}\left\langle k_{f}l_{f}\left|j_{L}(qr)\right|n_{i}l_{i}\right\rangle _{(\textrm{NR})}^{2}\delta(E\ldots)\,,
\end{align}
with the summations over $j_{f}$ and $j_{i}$ being done by the identity
\begin{equation}
\sum_{j_{f}=|l_{f}-\nicefrac{1}{2}|}^{l_{f}+1/2}\sum_{j_{i}=|l_{i}-\nicefrac{1}{2}|}^{l_{i}+1/2}[j_{f}]^{2}[j_{i}]^{2}\left\{ \begin{array}{ccc}
l_{f} & j_{f} & \nicefrac{1}{2}\\
j_{i} & l_{i} & L
\end{array}\right\} ^{2}=2\,.
\end{equation}
Note that the multipole rank is renamed from $J$ to $L$ to reflect
the fact that the operator is purely spatial. 

If one uses the continuum wave function normalization $\langle k^{'}l^{'}m_{l}^{'}|klm_{l}\rangle=(2\pi)^{3}/k^{2}\delta(k-k^{'})\delta_{l^{'}l}\delta_{m_{l}^{'}m_{l}}$
and integrate out $k_{f}=\sqrt{2m_{e}\varepsilon_{f}}$ through the
energy conservation delta function $\delta(\nicefrac{k_{f}^{2}}{2m_{e}}-\varepsilon_{n_{i}l_{i}}-T)$,
where $\varepsilon_{n_{i}l_{i}}$ is the energy of the $n_{i}l_{i}$
shell, 

\begin{subequations}
\begin{align}
\mathcal{R}{}_{\textrm{SI}}^{(\textrm{ion})}(T,q) & =\sum_{n_{i}l_{i}}\sum_{l_{f}}\sum_{L=0}\frac{2m_{e}k_{f}}{(2\pi)^{3}}[l_{f}]^{2}[l_{i}]^{2}[L]^{2}\left(\begin{array}{ccc}
l_{f} & L & l_{i}\\
0 & 0 & 0
\end{array}\right)^{2}\left\langle k_{f}l_{f}\left|j_{J}(qr)\right|n_{i}l_{i}\right\rangle _{(\textrm{NR})}^{2}\delta(E\ldots)\,,\\
 & =\sum_{n_{i}l_{i}}\frac{H(T+\varepsilon_{n_{i}l_{i}})}{4(T+\varepsilon_{n_{i}l_{i}})}\left|f_{\textrm{ion}}^{n_{i}l_{i}}(T,q)\right|^{2}\,.
\end{align}
\end{subequations}
This equation connects our definition of the SI response function
to the conventional atomic ionization form factor $f_{\textrm{ion}}^{n_{i}l_{i}}(T,q)$~\citep{Essig:2011nj,Essig:2012yx,Essig:2017kqs},
which is based on a nonrelativistic, independent particle approximation.
The Heaviside function $H(T+\varepsilon_{n_{i}l_{i}})$ is to ensure
the energy transfer is big enough to ionize a $n_{i}l_{i}$-shell
electron.

The SD response function can be worked out similarly with slightly
more cumbersome algebra. There are two key steps. One is the identity
that furnishes the summation over $j_{f}$ and $j_{i}$: 
\begin{equation}
\sum_{j_{f}=|l_{f}-\nicefrac{1}{2}|}^{l_{f}+1/2}\sum_{j_{i}=|l_{i}-\nicefrac{1}{2}|}^{l_{i}+1/2}[j_{f}]^{2}[j_{i}]^{2}[L]^{2}[1]^{2}\left\{ \begin{array}{ccc}
l_{f} & l_{i} & L\\
\nicefrac{1}{2} & \nicefrac{1}{2} & 1\\
j_{f} & j_{i} & J
\end{array}\right\} \left\{ \begin{array}{ccc}
l_{f} & l_{i} & L^{'}\\
\nicefrac{1}{2} & \nicefrac{1}{2} & 1\\
j_{f} & j_{i} & J
\end{array}\right\} =\delta_{LL'}\,.
\end{equation}
The other is re-organizing the summation over the rank of multipole
$J$ to the one over the rank of orbital angular momentum $L$. Among
the three types of operators involved, $\hat{\Sigma}_{J}$ has different
parity than $\hat{\Sigma}_{J}^{'}$ and $\hat{\Sigma}_{J}^{''}$.
For $\hat{\Sigma}_{J}$, the orbital rank $L=J$; for $\hat{\Sigma}_{J}^{'}$
and $\hat{\Sigma}_{J}^{''}$, the orbital rank can be $L=J\pm1$,
so we call them $\hat{\Sigma}_{J\pm}^{'}$ and $\hat{\Sigma}_{J\pm}^{''}$.
As the above 9-$j$ orthogonality shows, there is no mixing between
operators of different orbital rank. Therefore, for a given $L$,
the SD response function contains contributions from $|\Sigma_{L}|^{2}$,
$|\Sigma_{(L-1)+}^{'}|^{2}$, $|\Sigma_{(L-1)+}^{''}|^{2}$, $|\Sigma_{(L+1)-}^{'}|^{2}$,
and $|\Sigma_{(L+1)-}^{''}|^{2}$ with a requirement that $J\ge1$
for $\hat{\Sigma}_{J}$ and $\hat{\Sigma}_{J}^{'}$ and $J\ge0$ for
$\hat{\Sigma}_{J}^{''}$ . Eventually, we can reduce the SD response
function in the NR-IPA scheme as 

\begin{subequations}
\begin{align}
\mathcal{R}{}_{\textrm{SD}}^{(\textrm{ion})}(T,q)= & \sum_{k_{f}l_{f}}\sum_{n_{i}l_{i}}2[l_{f}]^{2}[l_{i}]^{2}\left\{ [1]^{2}\left(\begin{array}{ccc}
l_{f} & 0 & l_{i}\\
0 & 0 & 0
\end{array}\right)^{2}+\sum_{L=1}([L]^{2}+[L-1]^{2}+[L+1]^{2})\left(\begin{array}{ccc}
l_{f} & L & l_{i}\\
0 & 0 & 0
\end{array}\right)^{2}\right\} \nonumber \\
 & \times\left\langle k_{f}l_{f}\left|j_{L}(qr)\right|n_{i}l_{i}\right\rangle _{(\textrm{NR})}^{2}\delta(E\ldots)\nonumber \\
= & \sum_{k_{f}l_{f}}\sum_{n_{i}l_{i}}\sum_{L=0}6[l_{f}]^{2}[l_{i}]^{2}[L]^{2}\left(\begin{array}{ccc}
l_{f} & L & l_{i}\\
0 & 0 & 0
\end{array}\right)^{2}\left\langle k_{f}l_{f}\left|j_{L}(qr)\right|n_{i}l_{i}\right\rangle _{(\textrm{NR})}^{2}\delta(E\ldots)\\
= & 3\mathcal{R}{}_{\textrm{SI}}^{(\textrm{ion})}(T,q)\,,
\end{align}
\end{subequations}
and prove explicitly the scaling relation $\xi^{(\textrm{NR})}=3$. 

We note that the scaling relation $\xi^{(\textrm{NR})}=3$ can be
derived rather straightforwardly starting from the basis where the
spatial and spin part of the single particle wave functions are decoupled
and factorized. However, the lengthy derivation presented here should
help readers appreciate the differences between a truly many-body
and relativistic calculation from the ones based on nonrelativistic,
independent particle approximations. 
\end{document}